# Low Temperature Field-Effect in Crystalline Organic Material

V. Y. Butko[*†], J. C. Lashley[*] and A. P. Ramirez[‡,¶]

[*]*Los Alamos National Laboratory, Los Alamos, New Mexico, USA*
[‡]*Bell Laboratories, Lucent Technologies, 600 Mountain Avenue, Murray Hill, NY, USA*
[¶]*Columbia University, New York, NY*

Molecular organic materials offer the promise of novel electronic devices but also present challenges for understanding charge transport in narrow band systems. Low temperature studies elucidate fundamental transport processes. We report the lowest temperature field effect transport results on a crystalline oligomeric organic material, rubrene. We find field effect switching with on-off ratio up to $10^7$ at temperatures down to 10 K. Gated transport shows a factor of ~10 suppression of the thermal activation energy in 10-50 K range and nearly temperature independent resistivity below 10 K.

Interest in organic devices stems from their mechanical flexibility, their potential for interfacing to biological systems, and their ease of processing over large areas. [1-4]. Work on single-molecule devices also motivates the need for scalable approaches to integrated molecular electronics [5, 6] Unlike the development of inorganic semiconductor devices, applications for organics are well ahead of fundamental understanding, especially concerning scattering and trapping mechanisms and their chemical and morphological origins.

One of the key questions for organic semiconductors is what fraction of charge injected by gate bias in a field effect transistor (FET) configuration is either itinerant or localized, the latter due to deep-level trapping. The most stringent test of localization is to cool such a device to very low temperatures where the mobility edge can be probed without the complication of thermal activation. Low-temperature time-of-flight experiments on high-quality single crystals of naphthalene and anthracene suggest that band transport with mobility values, $\mu > 100$, is attainable [7]. In this work we present the first low-temperature measurements on field effect devices made from a crystalline oligomeric organic material, rubrene.

The first organic single crystal FETs made from rubrene, pentacene, and tetracene had carrier mobility ($\mu$) $\leq 1$ cm$^2$/(Vs) [8-11]. All of these devices demonstrated thermally activated resistivity, $\rho = \rho_0 \exp(E_a/k_B T)$, where $E_a$ is an activation energy, below 270 K. Recent work [12, 13] on rubrene single crystal FETs shows $\mu$ significantly higher than 1 cm$^2$/Vs in the temperature range 300K - 100K but, below 180 K, charge transport is still semiconducting-like with $E_a \sim 70$ meV. Rubrene ($C_{42}H_{28}$) is a molecule comprised of tetracene with four additional phenyl rings attached at the center positions. For crystal synthesis, we used a special batch of high-purity rubrene powder obtained from Aldrich. Crystal growth was achieved by successive runs using the horizontal physical vapor transport method of Laudise et. al. [14] using ultra high purity argon, at a flow rate of 0.5 ml/min. The source temperature was ramped slowly over a time period varying between 6 hours to two days, to the melting point 317 C. At the end of each run, the single crystals were visually examined for low mosaic spread by microscopy in cross-polarizers and the best crystals were recycled three times to further improve purity and crystallinity.

FETs were fabricated in a manner similar to our previous work using colloidal graphite source-drain contacts, parylene gate barriers, and silver-paste gate electrodes [9, 11]. Measurements were also performed as in our previous work [9, 11]. The leakage current between the gate electrode and ground was ~$10^{-14}$-$10^{-13}$A over most of the gate-source voltage ($V_{gs}$) range, and never exceeded $3\times10^{-12}$A. Heat sinking of the leads in the cryostat was achieved by either mechanical clamping to a cooled sapphire block in a vacuum of $10^{-5}$ Torr or with 10 Torr pressure $^4$He exchange gas. Additional thermal sinking is achieved by the three Au leads (2 cm by 50 μm). Thermal cooling between the sample and gate is limited by the parylene layer (100μm × 100μm) and is greater than $10^{-4}$ W/K, assuming a parylene thermal conductivity similar to other polymers ($\geq$ 0.01 W/(Km) [15]). At the highest power generated in our sample, we estimate temperature measurement error should not exceed ~2 K at the lowest temperatures. We emphasize that the above considerations address the rubrene lattice temperature and can only be used to bound an estimate of possible hot electron effects, to be discussed below.

Fig.1 demonstrates hole-injecting FET current-voltage (I-V) characteristics for a representative device (#6) at both 300K and at 10K. We find at 300K, linear behavior at small drain-source voltage ($V_{ds}$) followed by saturation at higher $V_{ds}$, behavior similar to that seen by other workers [12, 13]. (Sample geometries and room temperature mobilities for our devices are listed in Table 1.) We also find FET-like behavior at 10K in most samples. This behavior is qualitatively different in detail from that at 300K. In particular, the low-$V_{ds}$ behavior demonstrates a voltage threshold and does not fit the usual transport models, e.g. $I_d$ varying as V (ohmic), $V^2$ (space charge limited current), or $\exp(V^{1/2})$ (Frenkel-Poole emission). We suspect that the strong super-linear dependence of $I_d$ on $V_{ds}$ ($< V_{gs}$) is due to a combination of 1) contact barrier energy distribution, and 2) $V_{ds}$-stimulated shallow trap emission. It is also important to account for the role of contact potentials [3, 16] in the high-voltage charge transport. Previous work on rubrene FETs demonstrates that four- and two-terminal measurements yield similar results [8]. In the present experiments, both at 300K and 10K (fig. 1), the dependence of $I_d$ on $V_{gs}$ is much stronger than on $V_{ds}$ in the high-voltage limit. This behavior, and the effective saturation of $I_d$ on $V_{ds}$ at $V_{ds} \geq V_{gs}$ implies that contact potentials do not dominate high voltage transport.

Activation energy measurements performed at different values of $V_{gs}$, as shown in fig. 2, probe the density of deep traps as the Fermi level approaches the valence band [9, 11]. For instance, a gate voltage of –50V decreases $E_a$ from ~0.15 eV to 25 meV in the temperature range 270-120K. This latter value of $E_a$ is almost 3 times less than reported for rubrene in the temperature range 180-100 K [12] and almost 6 times less than in our pentacene FETs [9]. A smaller $E_a$ implies a significantly lower density of deep charge traps in these crystals, and subsequent higher channel conductivity at low temperatures, compared to previous published studies. We associate this higher conductivity with higher purity of our starting materials for crystal growth.

In Fig.3, we show $I_d$ versus temperature for three different rubrene FETs with data extending to lower temperatures than shown in fig. 2. Different values for $E_a$ can result from differences in processing and associated differences in trap densities within the first monolayer of rubrene at the interface to the gate barrier. A common feature to all devices, however, is a marked change in activation energy from $E_a \sim 25$-53 meV, to $E_a \sim 2$-5 meV at temperatures below 50K. Such behavior is often seen in small band gap systems, such as $SmB_6$ and FeSi, and arises from shallow states which become observable after the majority conduction band becomes thermally depopulated [17, 18]. In our devices, parametric gate bias control allows us to probe such states by varying charge density while keeping temperature fixed, as we discuss next.

Fig. 4 shows the low temperature dependence of the channel resistivity of rubrene FET device #2 calculated from the measured $I_d$ under the assumption of 1 nm channel depth [19] at different values of $V_{gs}$ for $V_{ds}$ = -125V. (Similar data were obtained for two other samples). These data clearly demonstrate gate-electric-field-induced crossover from thermally activated to nearly temperature independent transport as the hole Fermi energy moves toward the valence band [9, 11]. As discussed above, the temperature dependence of the channel resistance at the low-$V_{gs}$ observed in fig. 4 (inset) is due to a modification of the density of trap states available for thermal excitation. Then, for increasing $V_{gs}$ above ~ -80V, $E_a$ falls below the thermal energy, and activated behavior is lost. Indeed, for the high voltages, the resistivity between 2K and 30K cannot be fit with a single $E_a$ value. The differential number of injected holes per unit area is given by $dN = dV_{gs}C/e$, where C is the capacitance per unit area of the gate-insulator and e is the elementary charge. Therefore the effective hole mobility ($\mu_{eff}$) is a function of $V_{gs}$ and can be calculated from the measured $I_d(V_{gs})$ dependence (see inset of the fig.1): $\mu_{eff} = (dI_d/dV_{gs})Ld_{par}/(ZV_{ds}\varepsilon\varepsilon_0)$. Here L is drain-source contact separation, $d_{par}$ is parylene thickness, Z is the contact width, $\varepsilon = 2.65$ is the parylene dielectric constant, and $\varepsilon_0$ is the permittivity in vacuum. The mobility calculated for device #6 (fig.4, inset) displays three different regimes of the dependence of $\mu_{eff}$ on $V_{gs}$: 1) subthreshold behavior below 35V; 2) exponential increase of thermally activated carriers for 35V < $V_{gs}$ < 65V and; 3) $\mu_{eff} \propto V_{gs}^{m}$, above 65V, with m ~ 6.5. (Two other samples exhibited this behavior). We discuss this third region below.

We consider, below, two distinct transport scenarios that might explain the behavior at low temperature and high voltage. In the first scenario, we assume that all holes injected at the highest values of $V_{gs}$ are in the valence band and free. Under this assumption, one can consider the weak temperature dependence in fig. 4 as an approach to degeneracy. The density of free holes in the active channel is thus the differential amount injected by the gate electrode between -86 and -126V. We find the areal density of free holes of $\sim 5 \times 10^{11}$ cm$^{-2}$. Assuming a channel depth of 1 nm [19], the volume density of free holes is thus $5 \times 10^{18}$ cm$^{-3}$ (the rubrene molecular density is $1.5 \times 10^{21}$ cm$^{-3}$). The behavior of two-dimensional (2D) fermion gases in high-μ Si MOSFETs is known to be metallic-like ($d\rho/dT > 0$) at these areal densities. However, such behavior is only observed for $\mu > 4 \times 10^4$ cm$^2$/Vs [20], whereas for $\mu_{eff} \sim 0.5$ (the highest value obtained here - fig. 4 (inset)), metallic behavior has never been observed. Both insulator-metal (Ioffe-Regel) [21, 22] and insulator-superconductor [23] transition 2D criteria also classify our system as non-metallic. Even if our present system were three-dimensional, the observed value of $\mu_{eff}$ is too small for metallicity. In bulk P-doped silicon, for example, the limiting low temperature mobility is about 100 $cm^2/Vs$ at the density where the system crosses over from a semiconductor to a degenerate gas [24]. Therefore, it is difficult to reconcile an assumption of free holes with the observed low $\mu_{eff}$.

The second scenario incorporates the breakdown of the thermal activation model and a low $\mu_{eff}$ at low temperatures and invokes $V_{ds}$-induced trap emission followed by hopping, quantum tunneling, or hot electron transport. In general, trapped holes can hop or tunnel in real space between traps. The high-$V_{ds}$ fields of ~$10^4$ V/cm are most likely large enough to generate emission from trap states that are active at the low temperatures, and both hopping and tunneling can be activated by the shift of the Fermi level to the valence band. Both tunneling and hopping in the gap require a small average distance between traps and therefore a high density of trap states, as is typically observed in molecular compounds in proximity to the valence band [2]. An estimate of this distance obtained from the density of trapped injected carriers in our systems is ~10 nm, making such a mechanism unlikely at present carrier densities. On the other hand, trapped holes can be accelerated into the valence band, providing hot carrier transport between trapping events. Assuming again a 10 nm trap separation, one finds final kinetic energies of order 100K before trapping events. The central question that remains concerns the nature of trap states from which emission occurs. These states pin the Fermi level at low temperatures and the presently accessible voltages. The rapid mobility increase with $V_{gs}$ suggests that such pinning is due to a sharp rise in the density of states near a mobility edge [25].

In conclusion, we have shown that low temperature charge transport in a rubrene FET exhibits temperature dependence that crosses over from activated at high temperature to almost temperature-independent at 10K. The electric field dependence at the lowest temperatures suggests that trapping dominates charge transport within a hot-electron framework. Further work is needed in improving crystal purity to reduce trap density, and in improving the gate dielectric to increase injected charge density.

We are especially grateful to D. Lang for several useful discussions. We also acknowledge helpful discussions with C. M. Varma, R. de Picciotto, C. Kloc, X. Chi, X. Gao, S. Trugman and G. Lawes. We acknowledge support from the Laboratory Directed Research and Development Program at Los Alamos National Laboratory and by the DOE Office of Basic Energy Science.

†On leave from Ioffe Physical Technical Institute, Russian Academy of Science, Russia

### References

[1] P. Peumans, S. Uchida and S. R. Forrest, Nature **425**, 158 (2003).
[2] A. R. Volkel, R. A. Street and D. Knipp, Physical Review B **66**, 195336 (2002).
[3] I. H. Campbell and D. L. Smith, Solid State Physics **55**, 1 (2001).
[4] S. F. Nelson, Y. Y. Lin, D. J. Gundlach, et al., Applied physics letters **72**, 1854 (1998).
[5] C. Joachim, J. K. Gimzewski and A. Aviram, Nature **408**, 541 (2000).
[6] A. Nitzan and M. A. Ratner, Science **300**, 1384 (2003).
[7] W. Warta and N. Karl, Physical Review B **32**, 1172 (1985).
[8] V. Podzorov, V. M. Pudalov and M. E. Gershenson, Applied physics letters **82**, 1739 (2003).
[9] V. Y. Butko, X. Chi, D. V. Lang, et al., Appl. Phys. Lett. **83**, 4773 (2003).
[10] R. W. I. d. Boer, T. M. Klapwijk and A. F. Morpurgo, Appl. Phys. Lett. **83**, 4345 (2003).
[11] V. Y. Butko, X. Chi and A. P. Ramirez, Solid State Communications **128**, 431 (2003).
[12] V. Podzorov, E. Menard, A. Borissov, et al., cond-mat/0403575 (2004).
[13] R. W. I. d. Boer, M. E. Gershenson, A. F. Morpurgo, et al., cond-mat/0404100 (2004).
[14] R. A. Laudise, C. Kloc, P. G. Simpkins, et al., Journal of crystal growth **187**, 449 (1998).
[15] D. T. Morelli, J. Heremans, M. Sakamoto, et al., Phys. Rev. Lett. **57**, 869 (1986).
[16] A. Kahn, N. Koch and W. Y. Gao, Journal of Polymer Science, Part B (Polymer Physics) **41**, 2529 (2003).
[17] J. C. Cooley, M. C. Aronson, Z. Fisk, et al., Physical Review Letters **74**, 1629 (1995).
[18] S. Paschen, E. Felder, M. A. Chernikov, et al., Phys. Rev. B **56**, 12916 (1997).
[19] G. Horowitz, Advanced Functional Materials **13**, 53 (2003).
[20] E. Abrahams, S. V. Kravchenko and M. P. Sarachik, Reviews of modern physics **73**, 251 (2001).
[21] A. F. Ioffe and A. R. Regel, Prog. Semicond. **4**, 237 (1960).
[22] M. R. Graham, C. J. Adkins, H. Behar, et al., Journal of Physics: Condensed Matter **10**, 809 (1998).
[23] D. B. Haviland, Y. Liu and A. M. Goldman, Physical Review Letters **62**, 2180 (1989).

[24] G. L. Pearson and J. Bardeen, Phys. Rev. B **75**, 865 (1949).

[25] D. V. Lang, X. Chi, T. Siegrist, et al., cond-mat/0312722 (2003).

**Table 1.**

| Sample | L(μm) ± 20% | Z(μm) ±20% | $d_{par}$(μm) ±25% | μ($cm^2$/Vs) ±30% |
|---|---|---|---|---|
| #1 | 170 | 200 | 0.8 | 5 |
| #2 | 100 | 110 | 1.25 | 6 |
| #4 | 150 | 160 | 0.5 | 12 (285 K) |
| #6 | 100 | 150 | 1.25 | 5 |
| #12 | 100 | 100 | 1 | 2.5 |
| #18 | 150 | 250 | 0.8 | 3 |

**Figure 1**. The main parts show room temperature characteristics of the same rubrene single crystal FET at 300 K and 10 K. In the inset to fig. 1(a) is plotted $I_d$ versus $V_{gs}$ for the sample #4 with $V_{ds}$ = -51 V, # 6 with $V_{ds}$ = -61 V and # 2 with $V_{ds}$ = -61 V. In the inset to fig. 1(b) is plotted $I_d$ versus $V_{gs}$ for the sample #12 with $V_{ds}$ = -85 V, # 6 with $V_{ds}$ = -130 V and # 18 with $V_{ds}$ = -70 V.

**Figure 2**. Drain current temperature dependence of rubrene single crystal FET at different gate-source voltages. Inset: Dependence of the thermal activation energy on $V_{gs}$.

**Figure 3**. Drain current temperature dependence in 3 different rubrene single crystal FETs.

**Figure 4**. Resistivity temperature dependence of rubrene single crystal FET #2 at different gate-source voltages in temperature range 30K-2K. In the inset dependence of the mobility on $V_{gs}$ in the sample #6 at 10 K is shown.

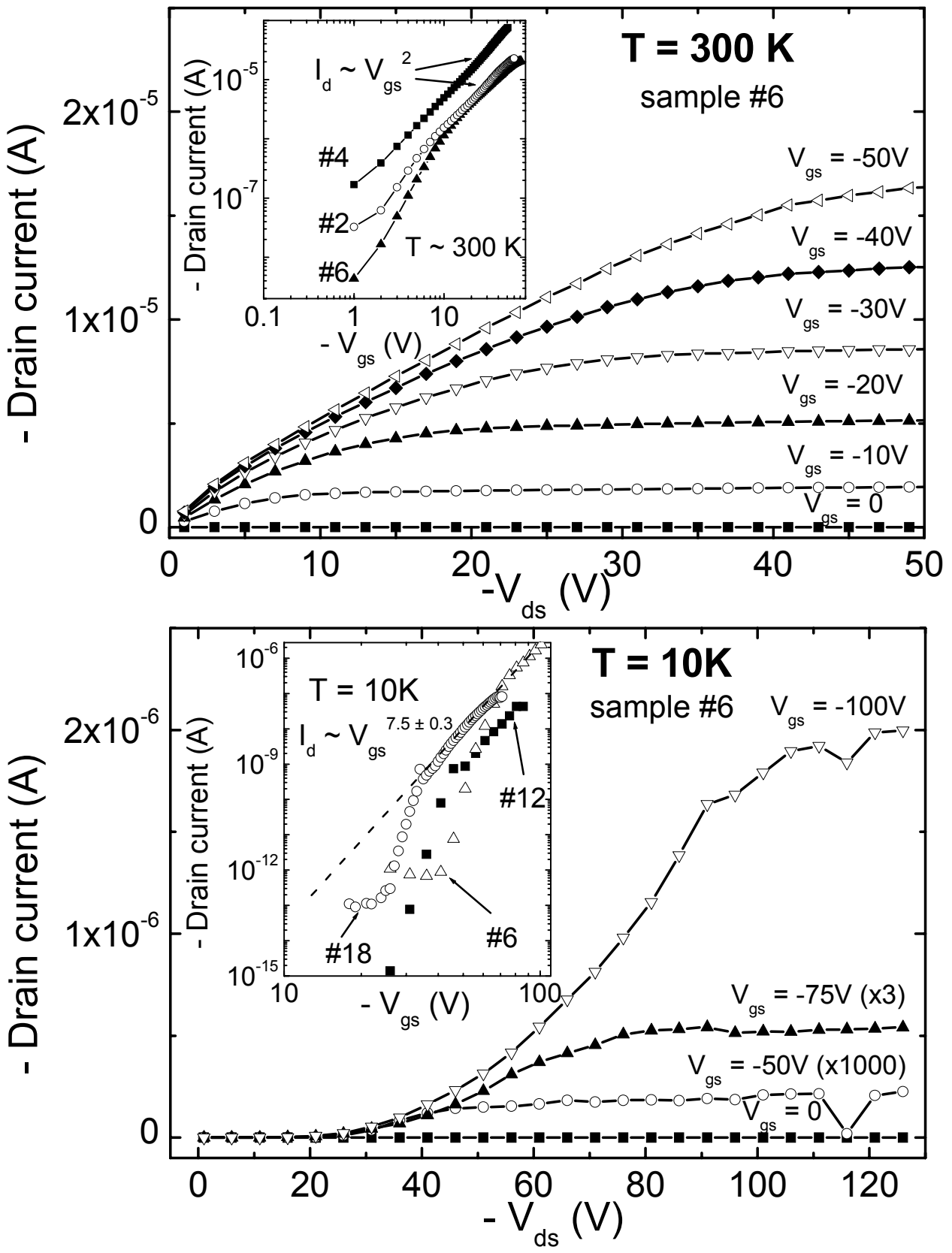

Figure 1.

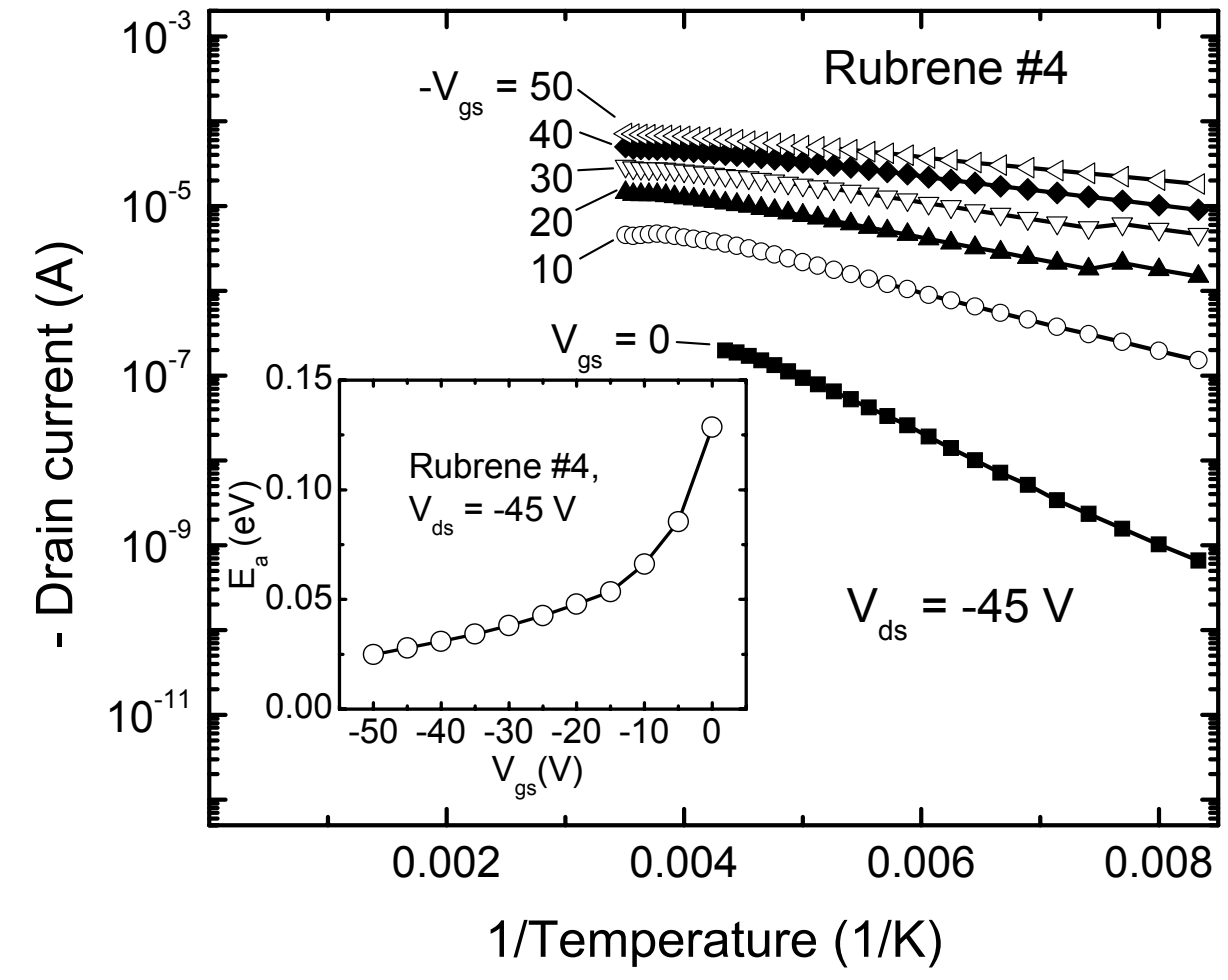

Figure 2.

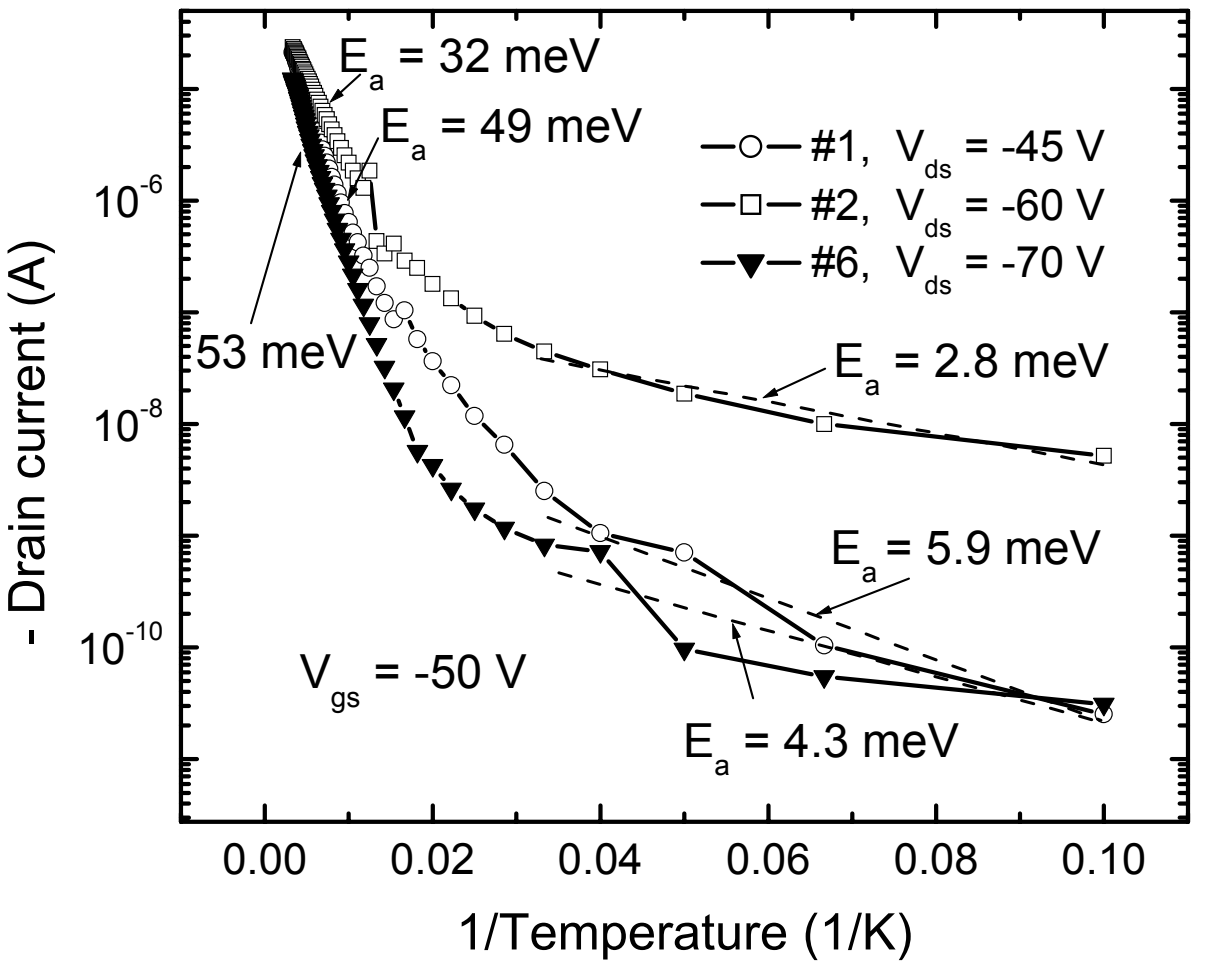

Figure 3.

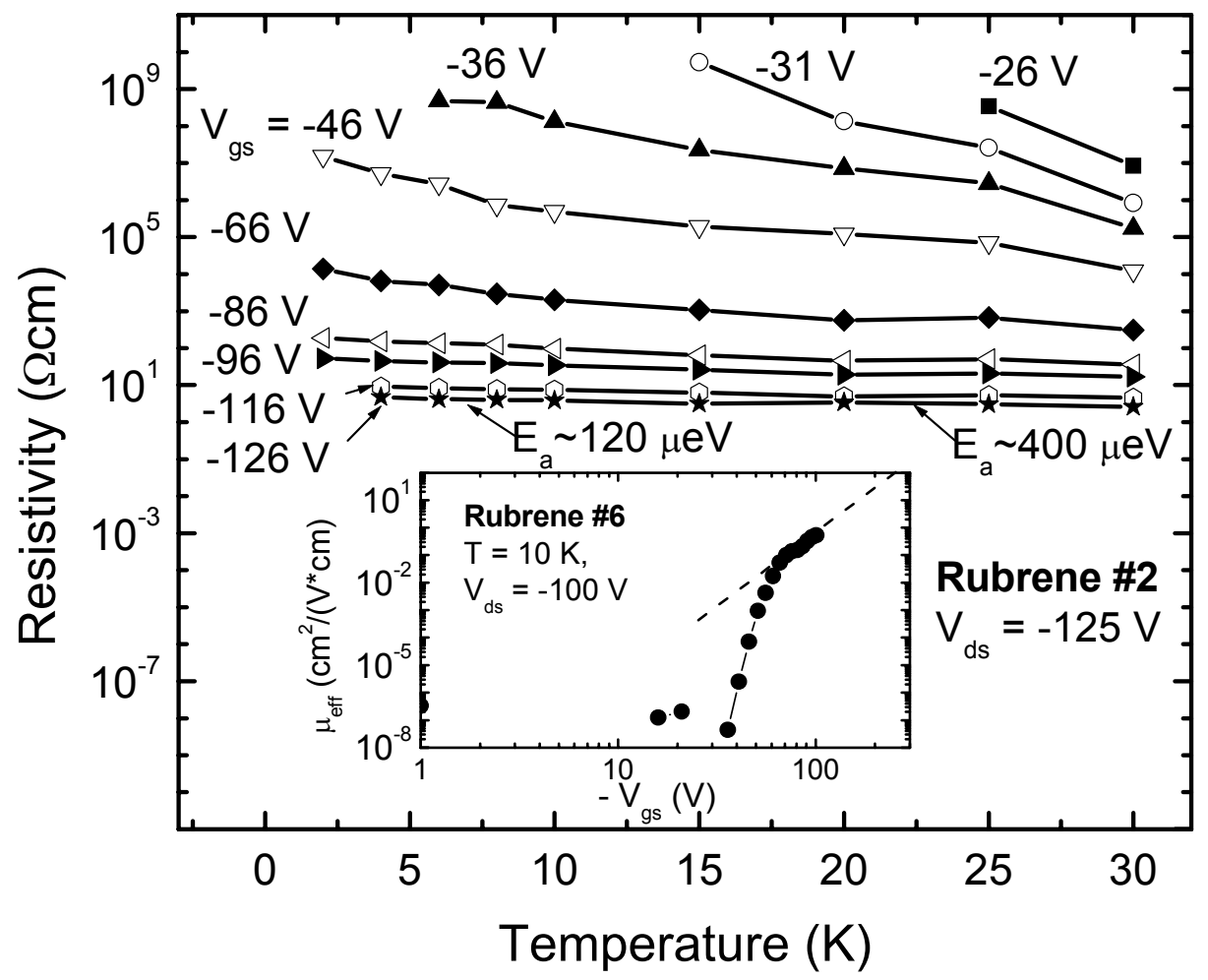

Figure 4.